**Non-adiabatic small polaron hopping in the n = 3 Ruddlesden-Popper compound $Ca_4Mn_3O_{10}$**

J. Lago[¶], P.D. Battle, M. J. Rosseinsky*[#] (Inorganic Chemistry Laboratory, University of Oxford)

A. I. Coldea, J. Singleton (Clarendon Laboratory, University of Oxford).

* Author for correspondence

[¶] Current address: Chemistry Department, University College London (UCL)

[#] Current address: Chemistry Department, University of Liverpool

**Abstract**

Magnetotransport properties of the compound $Ca_4Mn_3O_{10}$ are interpreted in terms of activated hopping of small magnetic polarons in the non-adiabatic regime. Polarons are most likely formed around $Mn^{3+}$ sites created by oxygen substoichiometry.

The application of an external field reduces the size of the magnetic contribution to the hopping barrier and thus produces an increase in the conductivity .We argue that the change in the effective activation energy around $T_N$ is due to the crossover to VRH conduction as antiferromagnetic order sets in.

## 1. Introduction:

Manganite perovskites of the type $RE_{1-x}M_xMnO_3$ (where RE = rare-earth and M = alkaline-earth, Pb) have been one of the hottest topics in condensed matter research since the discovery of the so-called colossal magnetoresistance (CMR) effect in the early nineties [1,2]. Although double-exchange (DE) between $Mn^{3+}$ ($d^4$) and $Mn^{4+}$ ($d^3$) was identified as early as 1951 as the basic mechanism of ferromagnetic (FM) coupling in these materials [3,4], it is now clear that it is the interplay of charge, orbital and spin degrees of freedom that have to be recalled to explain the rich magnetotransport behaviour encountered in the manganites [5].

In essence, within the DE mechanism, charge transport takes place through coherent hopping of the carrier between adjacent manganese sites, *i* and *j*. Due to a large on-site Hund's coupling, the effective hopping transfer integral is maximum when $S_i$ is parallel to $S_j$. As a result, a natural connection arises between electron hopping (and, hence, metallic conductivity) and ferromagnetic order. In this context, the drop in resistivity observed near the Curie temperature is qualitatively explained via a reduction by the applied field of the scattering arising from disorder of the localised spins. However, in contrast with what would be expected from this simple scenario, the experimental T-x phase diagrams of these materials show a pronounced electron-hole asymmetry and display a rich phenomenology of paramagnetic, antiferromagnetic (AFM), metallic FM, insulating FM and insulating charge-ordered (CO) phases. Early work mainly focused on the $x \sim \frac{1}{3}$ composition since it approximately represents the optimal electron density for ferromagnetism. More recently, however, attention has switched to doping levels for which competition between DE and AFM superexchange leads to separation into AFM and FM regions (i.e. boundary regions separating two competing states in phase space) as it is here that the largest CMR ratios are to be found, strongly linked to the presence of intrinsic inhomogeneities in the form of coexisting competing phases. In this picture, the insulator-to-metal transition appears percolative in nature: the FM metallic domains grow in size as $T \rightarrow T_c^+$ or a magnetic field is applied and finally percolate, short-circuiting the current and thus inducing the CMR effect [5-8].

A great deal of this renewed interest has been devoted to the electron-doped manganites, especially since the discovery that CMR also occurs in the $Mn^{4+}$ rich end of the $RE_{1-x}M_xMnO_3$ (M = Ca mainly) series for $x \sim x_c \cong 0.8 - 0.90$ [9,10]. In this regime, CMR is also linked to the competition between FM and CO AFM phases. The x = 1 end member of the series (y = 0, where y = 1-x is the electron-doping), CaMnO$_3$, is an insulating antiferromagnet with a G-type magnetic structure below T$_N$ ~ 123 K [11] due to the superexchange coupling between $t_{2g}$ electrons on neighbouring Mn$^{4+}$ sites. Substitution of Ca$^{2+}$ by a lanthanide ion causes these AFM interactions to compete with the FM that arises from the electron injection into the σ* $e_g$ band, leading to a series of different magnetic ground states as y increases. Interestingly, in contrast to earlier predictions [12], most experimental evidence indicate that a homogeneous long-range canting of the AFM background is unstable against segregation into a mixture of AFM and FM phases, even for the lowest electron-doping levels [13-15]. Thus, for y ≤ 0.045, doped electrons remain localized on single Mn sites leading to isolated Mn$^{3+}$ ions that coexist with the majority Mn$^{4+}$ sites [16,17]. The DE coupling to the neighbouring Mn$^{4+}$ ions distorts the magnetic environment locally creating a ferromagnetic cloud around itself. The result is an insulating regime in which small magnetoelastic polarons (local ferromagnetic regions involving a single Mn$^{3+}$ site) exist in an undistorted AFM lattice, so that the system retains the G-type magnetic structure. Above y ~ 0.045, this polaronic insulator becomes unstable as repulsion between $e_g$ electrons becomes sufficient to induce charge and/or orbital ordering. Thus, recent specific data for y = 0.06 are interpreted considering a long-wavelength spin density wave consisting of C-type AFM and FM components [16], the C structure corresponding to an orbital ordering in the *xy* planes. Further doping favours the C structure, which becomes prevailing for y ≥ 1-$x_c$. In this regime, the large MR ratios correspond to the melting of the CO state [18].

Oxygen deficiency in undoped CaMnO$_{3-\delta}$ (δ < 0.5) has the same effect of producing a mixture of Mn$^{3+}$ and Mn$^{4+}$ ions. However, in contrast to what happens with lanthanide substitution, the Mn$^{3+}$ cations are now in a square pyramidal rather than octahedral coordination. Magnetically, the system retains the G-type structure up to δ ~ 0.20, and transport remains thermally activated over the range 0.0 ≤ δ ≤ 0.33 [19]. As for the lanthanide-doped materials, charge ordering has also been found in oxygen-deficient CaMnO$_3$ [20]. Large MR has also recently been reported for δ ≤ 0.11 [21].

Here we present a detailed analysis of the thermopower and magnetotransport behaviour of Ca$_4$Mn$_3$O$_{10-\delta}$ (δ→ 0), the n = 3 member of the Ruddlesden-Popper (RP) series Ca$_{n+1}$Mn$_n$O$_{3n+1}$ of which CaMnO$_3$ can be regarded as the n = ∞ representative. RP phases (AO)(ABO$_3$)$_n$ [22] consist of perovskite slabs of *n* layers of corner-sharing BO$_6$ octahedra separated from neighbouring blocks by an AO rock-salt layer. The introduction of the rock-salt layer increases the number of bridging oxygens in the exchange pathway along the *z*-axis, thus causing a decrease in the magnitude of the exchange coupling in that direction. This produces an anisotropic reduction of the 3d bandwidth and hence modifies the magnetic and transport properties of these systems to an extent that depends on the value of *n*.

The magnetotransport properties of the low *n* RP manganites have been less studied than that of their $n = \infty$ counterparts. Activated conduction at high T has been reported for all undoped $Ca_{n+1}Mn_nO_{3n+1}$ materials [23], which, for n = 1, has been ascribed to small polaron hopping [24]. Large magnetoresistance ratios have been reported for the n = 2 lanthanide-doped materials [25,26]. As in the case of the $n = \infty$ perovskites, CMR in the mixed-valence bilayer manganites is not necessarily linked to the presence of PM→FM transition and, in fact, CMR effects have been reported without long-range ferromagnetism [27,28]. Again, experimental evidence suggests that phase separation is an intrinsic feature of these materials [5]. The same can be stated for the doped n = 1 manganites. In these systems, the quasi two-dimensional character precludes the onset of long-range ordered (*lro*) ferromagnetism and CMR. However, the magnetic phase diagram of $La_{1-x}Sr_{1+x}MnO_4$, with charge-ordered phases and a phase-coexistence regime [29,30], resembles that of $La_{1-x}Sr_xMnO_3$, indicating that similar physics is behind the phenomenology of these materials.

So far, only a handful of papers have been devoted to the n = 3 members of the RP series [31-33]. This is mainly due to the difficulty in producing mixed $Mn^{III}/Mn^{IV}$ materials by chemical substitution, which makes the system a poor candidate for CMR through the conventional DE mechanism. Attempts have been made to dope $Ca_4Mn_3O_{10}$ with $La^{3+}$ in order to induce mixed valence, but only small doping levels have been so far reported in bulk samples [32,34]. Recently, Yu and co-workers have shown that the reason for this rather low miscibility could be the instability of the as-obtained Pcab structure of the undoped $Ca_4Mn_3O_{10}$ [31] under the internal 'chemical' pressure originated by the difference in ionic radii between $Ca^{2+}$ and $La^{3+}$ [35,36]. As pressure is increased beyond ~ 19 MPa, the system undergoes a phase transition from the ambient-pressure orthorhombic structure to an undistorted tetragonal one (Space Group = I 4/mmm) [36].

Undoped $Ca_4Mn_3O_{10}$ orders antiferromagnetically at $T_N$ ~ 113 K (G-type magnetic structure), with a weak ferromagnetic (WFM) component that has been ascribed to a Dzyaloshinskii-Moriya term in the exchange Hamiltonian arising from the orthorhombic distortion of the crystal structure [33]. The WFM is hidden in zero field due to the AFM exchange along the *z*- axis but an applied magnetic field induces a spin-flop transition into a phase with a net magnetization. Magnetoresistance ratios up to 40 % have been reported for this material [37].

In this study of the electronic transport of $Ca_4Mn_3O_{10}$ in the vicinity of the AFM transition, we find activated transport over the entire measured T range (4.2 – 300 K) that we relate to the presence of small magnetoelastic polarons that originate from the localization of the small concentration of $e_g$ electrons that arise from oxygen deficiency. The large MR is explained in terms of the magnetic nature of the polarons.

## 2. Experiment

Powder samples of $Ca_4Mn_3O_{10-\delta}$ were prepared by the standard ceramic method. Details of the synthetic method have been described in earlier publications [31]. Purity of the products was established by means of X-ray and neutron diffraction measurements, via which it was found that the oxygen stoichiometry is not perfect, with $\delta < 0.04$ [31,33]. Resistivity measurements in zero and applied magnetic fields were carried out on small bars (2 x 2 x 5 mm$^3$) between 4.2 and 300 K using the four-point method with a direct current of less than 50 μA. The current-voltage dependence showed ohmic behaviour over the entire temperature range studied and the current was reversed during the measurements to eliminate the effects of thermal voltages. In order to minimize grain boundary effects, the sample was sintered at T slightly below its melting point (~1330 K). This way, we were able to obtain a density of ~ 84 % of the value determined from x-ray parameters.

A smaller bar cut from the same sample employed for the resistivity measurements was used to collect thermopower data. Data were collected between 70 and 340 K. The differential four-point steady-state method with separated measuring and power contacts was used to eliminate the thermal resistance. The temperature difference across the sample was kept constant between 0.3 and 1.0 K. Gold leads were used to detect the Seebeck voltage. No correction has been made on the collected data, for the measured values were several orders of magnitudes larger than any gold contribution ($S_{Au}$ = 1-2 μV/K).

## 3. Results

Figure 1 shows the thermal evolution of the resistivity of $Ca_4Mn_3O_{10}$ in zero field and in $B_{appl}$ = 14 T. It shows semiconducting behaviour over the range $T_N \pm 100$ K with no obvious singularity at $T_N$ = 113 K. However, a detailed examination of the data suggests two distinct regimes for temperatures above and below ~ 100 K. Figure 2 shows the thermal variation of the zero-field resistivity, ρ(T), for T > 100 K. In general, the observed exponential rise of ρ(T) on cooling reflects the activated nature of electronic transport in this regime. However, the Arrhenius plot $\ln\rho \propto T^{-1}$ fails to produce a straight line in any section of the current T range implying a thermal dependence of the pre-exponential factor and a departure, as expected, from a simple activation across a band gap characteristic of broad-band semiconductors. In analogy to previous studies on e-doped manganites, we have performed the analysis in terms of small polaron conduction, in which case the resistivity can be expressed by the general formula

$$\rho = \rho_0 \cdot T^\alpha \exp\left(-\frac{W}{kT}\right) \quad (1)$$

The dotted lines in Figure 2 correspond to the fit of the experimental data to a small polaron model in the adiabatic, α = 1 (W = 87.4(6) meV), and non-adiabatic, α = 3/2 (W = 93.6(4) meV) limits for T ≤ 200 K [38]. There is no significant difference between the two fits and, as reported for other oxide systems [39,40], it is difficult, if not impossible, to distinguish between the two regimes from this simple graphical analysis. Data collected in $B_{appl}$ = 14 T were also interpreted within a small polaron hopping model and a plot similar to that in Fig. 2 yields W = 81.9(5) and 88.0 (4) meV in the adiabatic and non-adiabatic regimes respectively. The application of a magnetic field seems, therefore, to lower the activation energy of the hopping process.

Below ~ 100 K, the resistivity (both in zero field and $B_{appl}$ = 14 T) is best described in terms of a general expression of Mott's variable range hopping (VRH) [41]

$$\rho = \rho_0 \exp\left(\left(\frac{T_0}{T}\right)^\nu\right) \quad (2)$$

where ν = 1/(d+1), d being the dimensionality of the system. Again, it is difficult to establish the magnitude of the exponent from a conventional $\ln\rho \propto T^{-\nu}$ plot, as a reasonably good fit is obtained for a wide range of values. Instead, we have calculated ν as the slope of the linear fit of the double logarithmic plot of the thermal variation of a local activation energy defined as $\Delta = -\frac{\partial \ln\rho}{\partial \ln T}$ [42]. These plots are displayed in Figure 3 for ZF and 14 T data. $T_0$ in Equation 2 is given by the relation $T_0 = \left(\frac{10^a}{\nu}\right)^{\frac{1}{\nu}}$, where $a$ is the intercept of the linear fit in Fig. 3. In this way, we obtain $\nu = 0.33(7)$ and $T_0 = 2.0 \times 10^6$ K in ZF data and $\nu = 0.32(9)$ and $T_0 = 2.3 \times 10^6$ K for the 14 T data. The inset in Fig. 3 shows the fit of the ZF experimental data to Greaves' generalized expression for the VRH model [43] in the temperature range $40 < T \leq 100$ K, using the calculated value of $\nu$.

A value of $\nu = 1/3$ is predicted for VRH transport in 2D. Thus, our analysis seems to indicate that the quasi-2D magnetic behaviour of $Ca_4Mn_3O_{10}$ [33] is also reproduced in the electronic transport. The fact that $\nu$ remains unchanged for the two sets of data further suggests that, as one might expect, the application of an external field does not alter the dimensionality of the transport. For a 2D VRH model, $T_0$ is expressed as [44]

$$T_0 = \frac{16\alpha^2}{kN(E_F)} \qquad (3)$$

where $\alpha$ is the inverse localization length of the hydrogenic wavefunction describing the localized state. Assuming small polaron conduction [45], we calculate a value of $\alpha$ of $\sim 6.6 \times 10^7$ cm$^{-1}$, which, from Eq. 3 results in a value of the 2D density of states at the Fermi level $N(E_F) \sim 4 \times 10^{14}$ eV$^{-1}$ cm$^{-2}$.

The derived values of $T_0$ compare well with those reported for other transition metal oxides [46-49]. However, extrapolating the calculated 2D-$N(E_F)$ to 3D yields a value of $2.7 \times 10^{22}$ eV$^{-1}$ cm$^{-3}$, somewhat higher than the values encountered in doped semiconductors ($10^{18}$-$10^{20}$ eV$^{-1}$ cm$^{-3}$ [50]). A value of $\alpha^{-1} \sim 10^1$ Å has been used before as an estimate for the localization length of carriers in the VRH regime [48,51], which would result in a lower value of $N(E_F)$. This implies a significant smearing of the wave function. However, this might not be correct in this case, in which the existence of small polarons requires that the carrier be localized onto a single lattice site. In order to test our calculations above, we have estimated the density of states at the Fermi level from the concentration of carriers derived from the thermopower measurements (see below). Following Wagner's work on the VRH regime in manganite perovskites [51,52], we have used a simple free-electron model expression of $N(E_F)$ [53]

$$N(E_F) = \frac{m^*}{\hbar^2 \pi^2} \cdot k_F = \frac{m^*}{\hbar^2 \pi^2} \cdot (3\pi^2 n)^{\frac{1}{3}} \qquad (4)$$

where we have approximated $m^* \sim 10\ m_e$ [51,53], since strong electronic correlations are expected in this narrow-band material. Substituting $n = 5.0 \times 10^{20}$ cm$^{-3}$ (see below) into Eq. 4 yields $N(E_F) \sim 3.3 \times 10^{22}$ eV$^{-1}$ cm$^{-3}$, which is comparable to the value calculated above from $T_0$. Substituting the new value of $N(E_F)$ into Eq. 4 results in a polaron radius of $r_p = \alpha^{-1} \sim 1.4$ Å similar to what has been found in other small polaron systems [54-57]. We, therefore, conclude that the calculated value of $T_0$ is consistent with a small polaron hopping mechanism. As discussed below, the large $N(E_F)$ could be explained by the small width of the polaronic band that forms upon reduction of the lattice by the oxygen deficiency.

The temperature dependence of the Seebeck coefficient, S(T), is presented in Figure 4. As expected, S(T) is negative over the whole thermal range studied, indicating that the majority carriers in Ca$_4$Mn$_3$O$_{10}$ are electrons.

The large magnitude ($|S| \sim 300$ μVK$^{-1}$, almost T-independent, at high temperatures) and its increase on cooling are, typical of semiconducting behaviour, in agreement with resistivity results above. In a simple approximation, the temperature dependence of the thermopower can then be expressed as [41]

$$S(T) = \frac{k}{e} \cdot \left( \frac{E_s}{T} + \alpha_{TH} \right) \qquad (5)$$

where $E_s$ is the effective gap energy ($E_s = |E_F - E'|$, where $E'$ is the average energy of the carriers) and $\alpha_{TH}$ is a sample-dependent constant [58]. For $E_s$ independent of T, $S \propto T^{-1}$. The linear fit in the inset of Fig. 4 (in the temperature range 100 – 200 K) yields a value of $E_s \sim 43$ meV, significantly below the value of the activation energy derived from the resistivity measurements, which, as we discuss in the following section, is characteristic of small polaron hopping transport.

Further evidence of polaronic hopping conduction arises from the evolution of S(T) above $\sim$ 150 K. In the case of polaron transport, for a fixed concentration of carriers, the thermopower approaches saturation when the thermal energy becomes larger than $w$, the polaronic bandwidth [59,60]. When this occurs, S(T) adopts a T-independent value

$$S_0 = -\left| \frac{k}{e} \ln\left( \beta \left( \frac{1-c}{c} \right) \right) \right| + S_1 \qquad (6)$$

where β is the spin-degeneracy factor, $c$ is the fraction of carries per available hopping site and $S_1$ is a constant such that $S_1$ times $k$T is the kinetic energy of the carriers. In any case, $S_1$ is predicted to be much smaller than the first term in the right-hand side of Eq. 6 ($S_1 < 10$ $\mu VK^{-1}$ [61]) and, hence, we have simply approximated it to zero. β measures the spin orientations allowed per site in the hopping process and can take the values 1 and 2. β = 1, i.e. no spin-degeneracy, is expected for hops in the ferromagnetic regime, whereas twofold spin degeneracy, β = 2, is predicted for hopping in a paramagnet [62]. In the current case, the strong intra-atomic Hund's coupling determines that only one spin orientation is allowed while hopping, and hence β = 1. Using the value of $S_0$ ~295 $\mu VK^{-1}$ estimated from Fig. 4, we thus calculate $c$ = 0.031. From the interpretation below that small polarons in $Ca_4Mn_3O_{10-\delta}$ correspond to electrons attached to the $Mn^{3+}$ sites originated by the oxygen substoichiometry, one electron per $Mn^{3+}$ ion, we take $c$ to be the fraction of $Mn^{3+}$ sites in relation to the total number of manganese. The value $c$ = 0.031 thus corresponds to the formula $Ca_4(Mn^{3+})_{0.093}(Mn^{4+})_{2.907}O_{9.954}$. From this, we calculate the concentration of carriers $n$ = 5.0 x $10^{20}$ $cm^{-3}$ used in calculations above.

## 4. Discussion
### 4.1. Small polaron hopping vs. activation to a mobility edge

The results presented in the previous section show that electronic transport in $Ca_4Mn_3O_{10-\delta}$ is activated in the entire temperature range, which we have analysed in terms of small polaron hopping. *A priori*, one could argue that activated conduction could as well originate from disorder (Anderson localization) or, simply, be the result of activation over a d-band splitting. However, the values of the activation energy derived from resistivity and thermopower measurements rule out these possibilities.

Were the system fully oxygenated (i.e. a perfect $Mn^{IV}$ $d^3$ system), activation over a d-band splitting would not be at all surprising. Simple crystal-field considerations show that the three $d$ electrons of $Mn^{IV}$ in the octahedral environment of the perovskite blocks lie in a half–filled $t_{2g}$ band above which lies an empty $e_g$ band. The $t_{2g}$ electrons are localized by the strong on-site Coulomb interactions that split the band into two one-electron sub-bands separated by U, as indicated by the presence of antiferromagnetic ordering in this material. The $e_g$ band is itself split due to the strong intra-atomic Hund energy. Within this simple picture, electronic conduction for the stoichiometric material would, therefore correspond to the activation of electrons from the lower $t_{2g}$ sub-band to the 'parallel spin' $e_g$ sub-band. To our knowledge, there are no reports of the magnitude of this gap for $Ca_4Mn_3O_{10}$. However, from analogy to $CaMnO_3$ (energy gap, ε, ~ 3-4 eV [63-65]), an energy gap of the order of $10^0$ eV can be expected for the present system, considerably larger than $W \sim 0.1$ eV calculated from our resistivity measurements.

The negative sign of the Seebeck coefficient is interpreted by assuming that the carriers in this material are electrons produced by oxygen sub-stoichiometry. The two electrons left by each missing oxygen are located on two $Mn^{3+}$ ions adjacent to the empty oxygen site. The carriers occupy states that are pulled bellow of the $e_g$ band due to the reduced ligand-field experienced in the local pyramidal symmetry compared to the regular octahedral one. In this scenario, the disorder generated by the removal of the oxygen atoms could be responsible of the localization of the carriers (Anderson localization) so that conduction would occur via activation to a mobility edge. This, in fact, has been proposed by Neumeier and Cohn [15] to explain the transport properties of $CaMnO_{3-\delta}$ ($\delta \sim 0$), which they ascribe to impurity states associated with chemical defects. However, our thermopower measurements rule out this interpretation in the current system. If transport occurred via activation of carriers to a mobility edge, both the resistivity and thermopower energy gaps would represent the energy required to create a mobile carrier, i.e. $|E_F - E'|$, and, therefore, be equal, which is not what we observe. If, on the other hand, transport occurs via thermally activated hopping of small polarons, then $W > E_s$. In this case [65,66]

$$\rho(T) = \rho_0 \cdot T^\alpha \exp\left(-\frac{\varepsilon_0 + W_H - t}{kT}\right)$$
$$S(T) = \frac{k}{e}\left(\frac{\varepsilon_0}{T} + \alpha\right)$$
(7)

where $\varepsilon_0$ is the energy difference between identical lattice distortions with and without the carrier; in other words, $\varepsilon_0$ is the energy required to generate the carrier, i.e. $E_s = \varepsilon_0 = |E_{F(0)} - E'|$ ($E_{F(0)}$ is the Fermi level of the undoped material). $W_H$ is the energy required to produce the equivalency of the two neighbouring sites in the hopping and $t$ is the transfer integral. In the case of thermopower measurements, hops occur between sites that have already been thermally activated and thus there is no such term as $W_H$ in $E_s$. We obtain $W \sim 0.09$ eV for the resistivity energy gap, much larger than $E_s \sim 0.04$ eV from the thermopower data, which shows that semiconducting behaviour observed in $Ca_4Mn_3O_{10}$ originates from small polaron hopping and not from activation to a mobility edge.

**4.2. Adiabatic vs. non-adiabatic hopping**

Earlier in the paper we mentioned that, in many cases, plots such as that in Figure 2 do not allow discrimination between the adiabatic and non-adiabatic regimes of small polaron hopping. This can be done, however, by a closer look at the Emin-Holstein theory [67,68]. In a small polaron system, electronic transport occurs when the electron, plus associated distortion, is able to move to an adjacent equivalent site overcoming the energy barrier that separate the initial and final state. At low temperatures, $T \ll \theta_D/2$, the only possible mechanism is quantum tunnelling. As temperature is raised, however, the width of the polaronic band decreases, tunnelling decays and, for $T \geq \theta/2$, thermally activated hopping begins to dominate. The hopping rate is maximized when the electron moves each time the local distortion pattern is the same for the two sites involved. In such a 'coincident event' [69], the electron transfers without introducing any further deformation and the hopping is termed adiabatic. The drift mobility is given by

$$\mu = \frac{3ea^2}{2} \frac{\omega_0}{2\pi kT} \exp\left(-\frac{W}{kT}\right) \qquad (8)$$

where $\omega_0$ is a characteristic phonon frequency and $a$ is the hopping distance.

In the non-adiabatic limit, the electron is no longer able to follow rapid fluctuations of the lattice and, hence, it does not respond quickly enough to the occurrence of a coincident event in order to overcome the energy barrier [70]. The mobility, reduced by a transfer probability $P \ll 1$ (compared to $P \sim 1$ in the adiabatic case) is now expressed as

$$\mu = \frac{3ea^2}{2\hbar} \frac{t^2}{kT} \left(\frac{\pi}{4WkT}\right)^{1/2} \exp\left(-\frac{W}{kT}\right) \qquad (9)$$

The probability of an electron being able to adjust to the atomic fluctuations is proportional to the transfer integral $t$. For a large $t$, the electron transfer is fast enough to follow the lattice fluctuations and the electronic transport occurs in the adiabatic regime. For small $t$ the hopping is non-adiabatic. Quantitatively, these considerations can be expressed in terms of the parameter $\eta_2 \equiv \frac{t^2}{\hbar\omega_0 \sqrt{kWT}}$ [70]; $\eta_2 > 1$ in the adiabatic limit, whereas for non-adiabatic transport $\eta_2 \ll 1$. In order to relate the nature of hopping to the value of the transfer integral this conditions can be re-written as

$$t >/\ll \left(\frac{2kTW}{\pi}\right)^{1/4} \left(\frac{\hbar\omega_0}{\pi}\right)^{1/2} \qquad (10)$$

In the case of $Ca_4Mn_3O_{10}$ we find $W \sim 0.09$ eV, in excellent agreement with a value of 96.8 meV reported by Greenblatt and co-workers [23] (compared to 130 meV reported earlier for this system [32], 70.7 meV for $CaMnO_3$ [21] and 100 meV for $Ca_2MnO_4$ [71]). We then assume a single frequency $\omega_0 \sim 10^{13}$ s$^{-1}$, characteristic of optical phonons in transition metal oxides, which results in a value of ~17 meV for the term on the right hand side of Eq. 10 at 100 K.

An estimate of the effective transfer integral $t$ can be obtained by considering the nature of the interactions between carriers and localized spins in this material. Each oxygen vacancy introduces two electrons into the otherwise Mn$^{IV}$ AFM insulator. The strong Hund's coupling (~ 3.4 eV in CaMnO$_3$ [64]) prevents the formation of a S = 1 state and the impurity electrons are parallel to the core $t_{2g}$ spins in localised states of mainly $d$ character and $e_g$ symmetry pulled below the conduction band ($e_g\uparrow$). Given the small concentration, a statistical distribution of Mn$^{3+}$ ions can be assumed, so that each Mn$^{3+}$ pair is surrounded by Mn$^{IV}$ ions. Electron transport occurs then via the hopping of the extra electrons along Mn$^{3+}$-O-Mn$^{IV}$ bonds (i.e. double exchange). The compound is thus a particular case of DE bound carriers in an AFM matrix, as studied by de Gennes [12]. Electron transfer can only occur between two adjacent sites and the relevant transfer integral is given by $t_{ij} = t_0 \cos(\theta_{ij}/2)$ [72] where $\theta_{ij}$ is the angle between the localised $t_{2g}$ spins. For an undistorted G-type AFM spin structure $t = 0$. However, in the case of Ca$_4$Mn$_3$O$_{10}$, the two magnetic sublattices are canted with a canting angle $\alpha \sim 0.13°$ [33], what gives raise to a small but finite value of $t$. The constant term $t_0$ is of the order of 0.1 eV [72-75] and $\cos(180-\alpha/2) \sim 10^{-3}$ for the observed small deviations from the perfect G-type arrangement. This leads to $t \sim 10^{-4}$ eV in Ca$_4$Mn$_3$O$_{10}$, two orders of magnitude smaller than the right-hand term in Eq. 10, from which it can be concluded that small polaron hopping occurs in the non-adiabatic regime in this material.

Note that this estimate of $t$ is strictly correct only for T ≤ T$_N$, since it is in this regime that $\alpha \sim 0.13°$. That is, $t \sim 10^{-4}$ represents a lower limit of the magnitude of the transfer integral above the transition. However, we believe that the presence of strong 2D AFM correlations at temperatures well above T$_N$ ensures that the picture of DE carriers in an AFM background remains the same at a local level, at least in the temperature range considered in this study. Thermal fluctuations, on the other hand, will play a more important role in the paramagnetic regime, and the decrease of the activation energy with increasing T (Figure 5, see next section) seems to indicate that they cause an increase of the average $\alpha$ above its value in the ordered structure.

### 4.3. Change in activation energy at T$_N$: onset of non-Arrhenius behaviour

Figure 5 shows the thermal dependence of the effective activation energy $\Delta^* = \partial\ln(\rho)/\partial(1/T)$ for the ZF and 14 T resistivities. Both curves show a sharp change in slope at ~ T$_N$; above this temperature the slope is negative, whereas below it the slope is positive. This change clearly shows that, despite the lack of a strong feature in the raw data at the onset of 3D-AFM ordering, electronic transport is being strongly affected by it. A change in activation energy has been found in previous studies of this material [23,32] (although in Ref [32] it is reported above $T_N$) and similar behaviour is also found in the n = 1 [23,24,71] and n = 2 [23] members of the $(CaMnO_3)_n(CaO)$ series. An increase in $E_A$ for T < 120 K has been, however, reported for stoichiometric $CaMnO_3$ [23].

An increase in the activation energy at the Néel temperature is expected for the hopping transport of a small polaron in an AFM system provided that it occurs via nearest-neighbours hops [76,77]. For a simple G-type antiferromagnet this means that all hops occur between non-degenerate states (as neighbouring sites belong to different sublattices) and thus the transfer rate is reduced by a magnetic factor from its value in a non-magnetic background and the polaron mobility decreases [77]. This behaviour has been observed in antiferromagnets such as NiO [78] and it could well be the reason behind the reported increase in activation energy in stoichiometric $CaMnO_{3.00}$. Oxygen-deficient $Ca_4Mn_3O_{10-\delta}$, however, differs significantly from NiO in that the relevant magnetic interactions are now ferromagnetic (through double-exchange) rather than antiferromagnetic.

The sharp change in activation at ~ 115 K most likely reflects the onset of non-Arrhenius temperature dependence as predicted theoretically for small polarons [79,80] when the multiphonon hopping processes that characterize the transport at high temperatures freeze out [69,80]. If only nearest-neighbour (nn) hops are considered, the lower limit for this high temperature regime is given by $T_0 \equiv \frac{\hbar\omega_0}{2k\sinh^{-1}\gamma}$ [39], where $\gamma = \frac{E_p}{\hbar\omega_0}$ is the small polaron coupling constant ($E_p = \frac{W_H}{2}$ is the polaron formation energy). This leads to $T_0$ = 108 K for $Ca_4Mn_3O_{10}$. In principle, hopping dies out below this point and tunnelling between extended states dominate. In practice, however, the width of the polaronic band is so small that the presence of potential fluctuations caused by impurities washes out the band entirely [81]. Transport occurs by electron transfer between localised states randomly distributed both in energy and space (i.e. VRH). In fact, we have already shown that the resistivity of $Ca_4Mn_3O_{10}$ at T < 100 K is best described by the VRH model.

The origin of this change in conduction mechanism is probably found in the DE link between the $e_g$-O-$e_g$ transfer integral and magnetic ordering of the $t_{2g}$ spins discussed above. As T is lowered, the strong 2D-AFM correlations that develop in the *ab* planes lower the transfer integral from the value in the paramagnetic regime. At T ~ $T_N$, the 3D-AFM ordering makes the activation energy so large for nearest-neighbour hops, that hopping further beyond nearest-neighbours start to be statistically significant, i.e. a VRH regime sets in. This effectively reduces the effect of AFM ordering on transport and, hence, $\Delta^*$ decreases.

A similar explanation has been proposed by Fisher for the change in $\partial S/\partial T$ they observe at $T_N$ for the n = 1, 2, ∞ members of $(CaMnO_3)_n(CaO)$ [82] and the antiferromagnet $BaCoS_2$ [83].

### 4.4 Field dependence of the electronic transport

The thermal evolution of both the raw resistivity and the effective activation energy in zero field and 14 T ($\Delta^*$ in 14 T is consistently lower than in zero-field in Fig. 5) clearly show that the application of an external magnetic field has a significant effect upon electronic transport over the entire temperature range. In Figure 6 we present the field dependence of MR = $(\rho(B)-\rho(0))/\rho(0)$ at different temperatures above and below $T_N$. In contrast to what is observed in other CMR systems, the parabolic dependence on applied field at both sides of the magnetic transition indicates that a similar, if not the same, mechanism is behind the large magnetoresistance in both the paramagnetic and antiferromagnetic regimes (MR ~ 40 % in 14 T at 61 K).

In our previous report on the physical properties of $Ca_4Mn_3O_{10}$ [37], we showed that the magnetoresistance is proportional to the squared magnetization at all measured temperatures (see Figure 7), which suggests that it results from the reduction by the applied field of the spin-dependent scattering of the carriers. Given that data were taken on powder samples of the compound, a spin-dependent mechanism of MR could, in principle, have an intrinsic as well as an extrinsic origin. In the latter, the $M^2$ dependence is associated with spin-polarised tunnelling between ferromagnetic grains. However, this *granular* mechanism is usually significant in the presence of magnetic order, quite the opposite of what is found in the current case. We thus believe that the observed MR is intrinsic to the system and related to the hopping of magnetic polarons.

The self-localization of a double-exchange carrier in a bipartite antiferromagnet was first considered by de Gennes [12], who showed that in the unperturbed spin structure it is energetically favourable for each localised carrier to build up a local spin distortion in its vicinity, in which it becomes further trapped. From analogy to lattice polarons, the quasi-particle thus produced (formed by the electron and a ferromagnetic cloud that surrounds it) is termed a magnetic polaron. The energy stabilization produced by the spin distortion acts as the binding for the polaron and transport occurs via activated hopping.

In the manganites, polarons of dual nature have been found (magnetoelastic polarons). In simple terms, the basic idea behind these is that the induced magnetic cloud around the localized carrier deepens the potential well in which it is trapped and thus extends the degree of localization. This results in a magnetic contribution to the hopping activation energy, which, because of the particular characteristics of double-exchange, depends on the misorientation between the localized spins at the neighbouring sites between the hopping process occurs. The application of an external field or the onset of a positive mean field (ferromagnet) aligns the local spins and thus reduces the magnitude of the activation barrier, which results in the drop of resistivity.

Quantitatively, a model that accounts for the MR in manganites based on activated hopping with additional barriers due to magnetic misalignment of the local moments has been developed by Wagner and co-workers, who expressed the new activation barrier as [51,52]

$$W_{ij}' = W_{ij} - \alpha (\vec{M}_i \cdot \vec{M}_j) \qquad (11)$$

where $W_{ij}$ is the non-magnetic hopping barrier and $M_i$ and $M_j$ are the local magnetization vectors at both sites of the hopping process. The effective barrier is therefore enhanced for a paramagnet or antiferromagnet but decreases for a ferromagnet. In the paramagnetic phase the average $\overline{(\vec{M}_i \cdot \vec{M}_j)}$ can be approximated by the squared Brillouin function B whereas below an ordering transition it is the Brillouin function itself that best describes the misorientation of adjacent magnetization vectors and thus the average above (see Refs. [51,52] for a formal development of the model). Given that the application of an external magnetic field does not affect the term $W_{ij}$, the drop in resistivity (i.e. the negative MR) is, in the limit $W'_{ij} \ll kT$, proportional to $\overline{(\vec{M}_i \cdot \vec{M}_j)}$ and, therefore, to $B^2$ and B above and below an ordering magnetic transition respectively.

Red solid lines in Figure 7 represent the attempts to fit the experimental MR vs. $B_{appl}$ data to the squared Brillouin function at different temperatures above and below $T_N$ for $Ca_4Mn_3O_{10}$ leaving $J$, the effective magnetic moment, as the only varying parameter. The agreement is excellent for $T > T_N$ but fails below the transition. Note that the fit to B for $T < T_N$ (not presented) yields an even worse fit to the experimental data than that to $B^2$ in Fig. 7.

The quality of the fits above the transition confirms the current interpretation of electronic transport in $Ca_4Mn_3O_{10}$. The derived values of $J$ (5.36 and 5.19 $\mu_B$ at 161 and 223 K respectively), slightly larger than expected for a ferromagnetically coupled $Mn^{3+}$-$Mn^{4+}$ pair (4.5 $\mu_B$), agrees with the idea that conduction involves the activated hopping of magnetic polarons formed by a single localised carrier and its induced cloud of ferromagnetically polarised neighbours. For the sake of comparison, an effective moment $J_{eff}$ = 6 $\mu_B$ has been reported for $La_{0.97}MnO_3$, for which the presence of small magnetic polarons is well established [84].

Below $T_N$, $J$ increases up to ~ 7.2 $\mu_B$ at 6 1K. However, the poor fit to the model prevents us from reaching any conclusions as to the effect of the 3D-AFM ordering in the size of the polarons. It is worth noting that in this temperature range a squared Brillouin function still provides the best, albeit deficient, description to the experimental data, in contrast to what is predicted by the model. This seems to indicate that the conduction in the ordered state does not differ significantly from that in the paramagnetic regime, which could be related to the fact that 2D-AFM correlated regions develop at temperatures quite above $T_N$ and transport occurs mainly in the *ab* planes (from the value of ν in Eq. 2). In this sense, it is interesting to mention that MR ∝ $M^2$ in the entire T range, pointing again towards a common origin for the effect.

Wagner's B-dependence of MR in the ordered state derives from the assumption that the applied field has no influence over the Weiss magnetization, which may not be the case in the present system, where we have shown [33] that a small net magnetization is induced by the application of an external field. In fact, our analysis of the high temperature resistivity is based on the assumption that DE occurs due to the canting of the magnetic sublattices rather than the existence of local ferrimagnetism as proposed by Neumeier for $CaMnO_3$ [15].

## 5. Conclusions

The results presented in this paper show that the activated nature of the electronic transport in $Ca_4Mn_3O_{10}$ results from the hopping of small polarons of dual magnetic-lattice character formed around $Mn^{3+}$ impurities left behind by the removal of oxygen. At high temperatures, hopping occurs between neighbouring sites whereas below ~ 100 K jumps beyond nearest neighbours yield a Mott's VRH-type resistivity. Above $T_N$, transport occurs non-adiabatically due to the small transfer integral.

The application of an external magnetic field produces a decrease in the activation energy thus giving rise to a large negative magnetoresistance. MR is proportional to the squared magnetization over the entire temperature range studied suggesting a spin-dependent scattering mechanism. Data above $T_N$ agree with a model by Wagner and co-workers in which the magnetic-dependent energy barrier is a function of the average misorientation of local magnetization vectors. However, this model predicts a linear dependence on B in the ordered magnetic state below $T_N$ in contrast with our findings. The reason for this disagreement is not yet clear. However, it could be related to the presence of a field-induced weak ferromagnetism in the G-type ordered $Ca_4Mn_3O_{10}$.

**Acknowledgements**

The authors, especially J L, are indebted to Dr. J. Cooper from the IRC for Superconductivity, University of Cambridge for the thermopower measurements and useful discussion. J L wants to thank the Basque Government-Eusko Jaurlaritza for a pre-doctoral research grant.

Reference List

1. Kusters,R.M., Singleton,J., Keen,D.A., Mcgreevy,R. & Hayes,W. *Phys. B* **155**, 362 (1989).

2. Von Helmolt,R., Wecker,J., Hozapfel,B., Schultz,L. & Samwer,K. *Phys. Rev. Lett.* **71**, 2331 (1993).

3. Zener,C. *Phys. Rev.* **81**, 440 (1951).

4. Zener,C. *Phys. Rev.* **82**, 403 (1951).

5. Dagotto,E., Hotta,H. & Moreo,A. *Phys. Rep.* **344**, 1 (2001).

6. Mayr,M. *et al. Phys. Rev. Lett.* **84**, 5568 (2000).

7. Burgy,J., Mayr,M., Martin-Mayor,V., Moreo,A. & Dagotto,E. *Phys. Rev. Lett.* **87**, 277202 (2001).

8. De Teresa,J.M. *et al. Phys. Rev. B* **65**, 100403 (2002).

9. Martin,C., Maignan,C., Damay,F., Hervieu,M. & Raveau,B. *J. Solid State Chem.* **134**, 198 (1997).

10. Martin,C., Maignan,C., Damay,F., Hervieu,M. & Raveau,B. *Phys. Rev. B* **60**, 12191 (1999).

11. MacChesney,J.B., Williams,H.J., Potter,J.F. & Sherwood,R.C. *Phys. Rev.* **164**, 779 (1967).

12. de Gennes,P.G. *Phys. Rev.* **118**, 141 (1960).

13. Maignan,C., Martin,C., Damay,F., Raveau,B. & Hejtmanek,J. *Phys. Rev. B* **58**, 2758 (1998).

14. Savosta,M.M. *et al. Phys. Rev. B* **62**, 9532 (2000).

15. Neumeier,J.J. & Cohn,J.L. *Phys. Rev. B* **61**, 14319 (2000).

16. Cornelius,A.L., Light,B. & Neumeier,J.J. cond-mat/0108239 . 2001.

17. Chen,Y.-R. & Allen,P. *Phys. Rev. B* **64**, 064401 (2001).

18. Chiba,H., Kikuchi,M., Kusaba,K., Muraoka,Y. & Syono,Y. *Solid State Commun.* **99**, 499 (1996).

19. Briatico,J. *et al. Phys. Rev. B* **53**, 14020 (1996).

20. Reller,A., Thomas,J.M. & Jefferson,D.A. *Proc. R. Soc. A* **394**, 223 (1984).

21. Zeng,Z., Greenblatt,M. & Croft,M. *Phys. Rev. B* **59**, 8784 (1999).

22. Ruddlesden,S.N. & Popper,P. *Acta Crystallogr.* **11**, (1958).

23. Fawcett,I.D., Sunstrom,J.E., Greenblatt,M., Croft,M. & Ramanujachary,K.V. *Chem. Mater.* **10**, 3643 (1998).

24. Jung,W.H. & Iguchi,E. *J. Phys. D: Appl. Phys.* **31**, 794 (1998).

25. Moritomo,Y., Asamitsu,A., Kuwahara,H. & Tokura,Y. *Nature* **380**, 141 (1997).

26. Asano,H., Hayakawa,J. & Matsui,M. *Appl. Phys. Lett.* **68**, 3638 (1996).

27. Battle,P.D. *et al. J. Phys. : Condens. Matter* **8**, L427 (1996).

28. Hur,N.H. *et al. Phys. Rev. B* **57**, 10740 (1998).

29. Bao,W., Chen,C.H., Carter,S.A. & Cheong,S.-W. *Solid State Commun.* **98**, 55 (1996).

30. Larochelle,S. *et al. Phys. Rev. Lett.* **87**, 095502 (2001).

31. Battle,P.D. *et al. Chem. Mater.* **10**, 658 (1998).

32. Witte,N.S., Goodman,P., Lincoln,F.J., March,R.H. & Kennedy,S.J. *Appl. Phys. Lett.* **72**, 853 (1998).

33. Lago,J., Battle,P.D. & Rosseinsky,M.J. *J. Phys. : Condens. Matter* **12**, 2505 (2000).

34. Asano,H., Hayakawa,J. & Matsui,M. *Appl. Phys. Lett.* **71**, 844 (1997).

35. Yu,R.C. *et al. J. Appl. Phys.* **90**, 6302 (2001).

36. Yu,R.C. *et al. J. Appl. Phys.* **91**, 6765 (2002).

37. Mihut,A.I. *et al. J. Phys. : Condens. Matter* **10**, L727 (1998).

38. Above 200 K, there seems to be a crossover to regime with lower activation energy W ~ 42 meV, as indicated by a change of slope in the semilog plot in Fig. 2. This could be an artificial artefact, as it has not been observed in previous studies, or indicate a new hopping regime in which 2D-AFM correlations play a much lesser role. In any case, we believe it does not affect the validity of our conclusions for the temperature range under study. It would be desirable, however to extend the current measurements to higher T.

39. Casado,J.M., Harding,J.H. & Hyland,G.J. *J. Phys. : Condens. Matter* **6**, 4685 (1994).

40. Jaime,M. *et al. Phys. Rev. Lett.* **78**, 951 (1997).

41. Mott,N.F. *J. Non-Cryst. Solids* **1**, 1 (1968).

42. Hill,R.M. *Phys. Stat. Solidi A* **35**, K29 (1976).

43. Greaves,G.N. *J. Non-Cryst. Solids* **11**, 427 (1973).

44. Bogolomov,V.N., Kudinov,E.K. & Firsov,Y.A. *Sov. Phys. -Solid State* **9**, 2502 (1988).

45. In this case, the polaron radius, i.e. the localization length of the quasiparticle, can be approximated (Ref. 44) by $r_p = 1/2(\pi/6)^{1/3}R$, where R is the average intersite separation. For $Ca_4Mn_3O_{10}$, R ~ 3.7 Å within the perovskite slabs in which electronic transport is most likely to occur. Hence, $r_p$ ~ 1.5 Å and $\alpha$ ~ 0.67 Å$^{-1}$.

46. Ziese,M. & Srinitiwarawong,C. *Phys. Rev. B* **58**, 11519 (1998).

47. Anil Kumar,P.S., Joy,P.A. & Date,S.K. *J. Phys. : Condens. Matter* **10**, L269 (1998).

48. Kastner,M.A. *et al. Phys. Rev. B* **37**, 111 (1988).

49. Fisher,B. *et al. Phys. Rev. B* **50**, 4118 (1994).

50. Mott,N.F. Conduction in Non-Crystalline Materials. Clarendon, Oxford (1979).

51. Wagner,P.H. *et al. Phys. Rev. B* **55**, 3699 (1997).

52. Wagner,P.H. *et al. Phys. Rev. Lett.* **81**, 3980 (1998).

53. Ashcroft,N.W. & Mermin,N.D. Solid State Physics. Saunders College, (1976).

54. Mollah,S., Som,K.K. & Bose,K. *Phys. Rev. B* **46**, 11075 (1992).

55. Doweidar,H., El-Damrawi,G.M. & Moustafa,Y.M. *J. Phys. : Condens. Matter* **6**, 8829 (1994).

56. Ghosh,A. *J. Phys. : Condens. Matter* **5**, 8749 (1993).

57. Chaterjee,S., Bhattacharya,S. & Chaudhuri,B.K. *J. Chem. Phys.* **108**, 2954 (1998).

58. Munakata,F., Matsuura,K., Kubo,K., Kawano,T. & Yamauchi,H. *Phys. Rev. B* **45**, 10604 (1992).

59. Fisher,B., Patlagan,L. & Reisner,G.M. *Phys. Rev. B* **54**, 9359 (1996).

60. Schmidbauer,E. *J. Solid State Chem.* **134**, 253 (1997).

61. Schmidbauer,E. *J. Phys. : Condens. Matter* **10**, 8279 (1998).

62. Chaikin,P.M. & Beni,G. *Phys. Rev. B* **13**, 647 (1979).

63. Zampieri,G. *et al. Phys. Rev. B* **58**, (1998).

64. Jung,J.H. *et al. Phys. Rev. B* **55**, 15489 (1997).

65. Jaime,M. *et al. Phys. Rev. B* **54**, 11914 (1996).

66. Wang,S., Li,K., Chen,Z. & Zhang,Y. *Phys. Rev. B* **61**, 575 (2000).

67. Holstein,T. *Ann. Phys.* **8**, 343 (1959).

68. Emin,D. & Holstein,T. *Ann. Phys.* **53**, 439 (1969).

69. Emin,D. *Phys. Today* **35**, 34 (1982).

70. Bottger,H. & Bryksin,V.V. Hopping Conduction in Solids. VCH, Weinheim (1985).

71. Yamashita,T. *et al. Phys. Rev. B* **53**, 14470 (1996).

72. Anderson,P.W. & Hasegawa,H. *Phys. Rev.* **100**, 675 (1955).

73. Boris,A.V. *et al. Phys. Rev. B* **59**, R697 (1999).

74. Yi,H. & Yu,J. *Phys. Rev. B* **58**, 11123 (1998).

75. Millis,A.J., Shraiman,B.I. & Mueller,R. *Phys. Rev. Lett.* **77**, 175 (1996).

76. Appel,J. *Phys. Rev.* **141**, 506 (1966).

77. Emin,D. & Liu,N.L.H. *Phys. Rev. B* **27**, 4788 (1983).

78. Keem,J.E., Honig,J.M. & Van Zandt,L.L. *Phil. Mag. B* **37**, 537 (1978).

79. Austin,I.G. & Mott,N.F. *Adv. Phys.* **18**, 41 (1969).

80. Emin,D. *Adv. Phys.* **24**, 305 (1975).

81. Bosman,A.J. & van Daal,H.J. *Adv. Phys.* **19**, 1 (1970).

82. Fisher,B., Patlagan,L., Reisner,G.M. & Knizhnik,A. *Phys. Rev. B* **61**, 470 (2000).

83. Fisher,B., Genossar,J., Patlagan,L., Reisner,G.M. & Knizhnik,A. *Phys. Rev. B* **59**, 8745 (1999).

84. de Brion,S., Chouteau,G. & Lejay,P. *Phys. B* **259-261**, 818 (1999).

**Figure Captions**

1. Temperature dependence of the resistivity of $Ca_4Mn_3O_{10}$ in zero-field and B = 14 T.

2. Adiabatic ($\alpha = 1$) and non-adiabatic ($\alpha = 3/2$) polaron hopping fits (red dotted lines) to the zero-field resistivity data above 100 K.

3. Temperature dependence of the local activation energy $\Delta = -\partial \ln\rho/\partial \ln T$, used to determined the exponent $\nu$ and $T_0$ in the general expression for VRH. **Inset:** Fit of the low temperature ZF resistivity to a VRH model in two dimensions.

4. Thermal evolution of the Seebeck coefficient for $Ca_4Mn_3O_{10}$. **Inset:** 1/T-dependence of S(T) as expected for small polaron hopping.

5. Thermal evolution of an effective activation energy $\Delta^* = \partial \ln\rho/\partial(1/T)$ for the zero-field (o) and B = 14 T (•) resistivities.

6. Magnetic field dependence of the magnetoresistance MR = $(\rho(0)-\rho(B))/\rho(0)$ for $Ca_4Mn_3O_{10}$. The solid red lines represent the fit to the squared Brillouin function leaving the effective moment $J_{eff}$ as the only varying parameter.

7. Linear relationship between MR and the squared experimental magnetization at different temperatures above and below $T_N$.

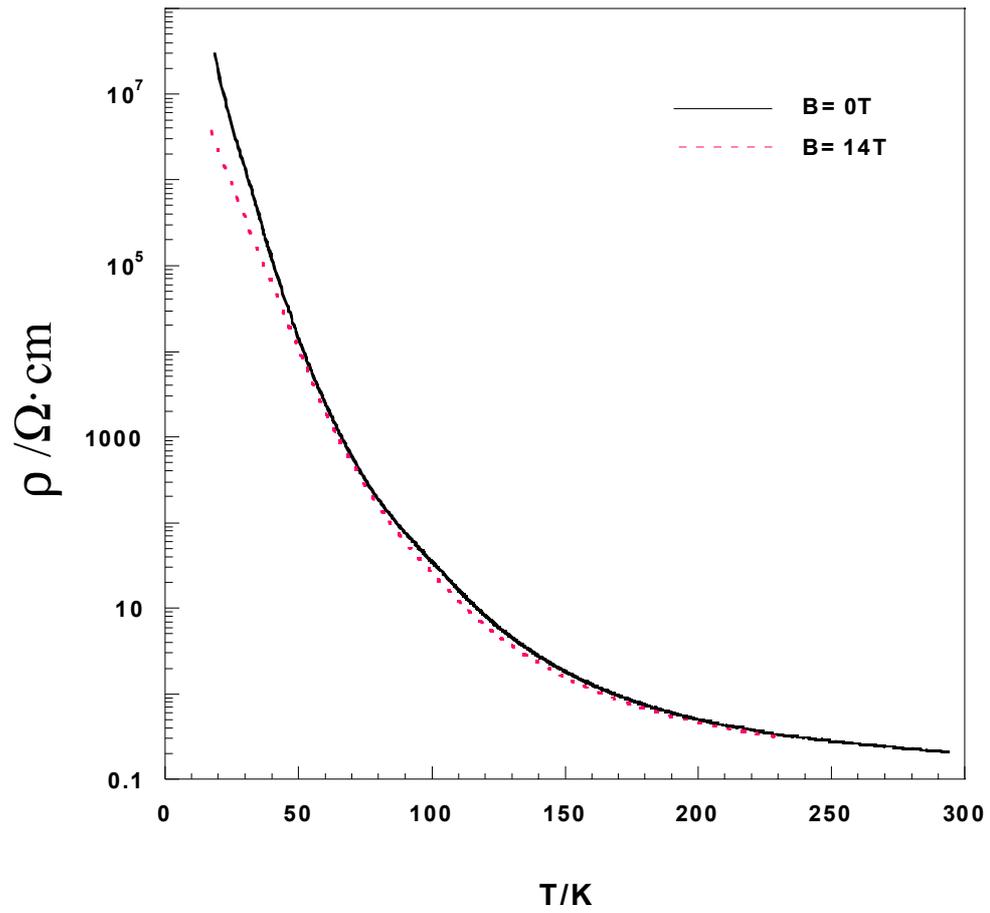

Figure 1 (Lago et al.)

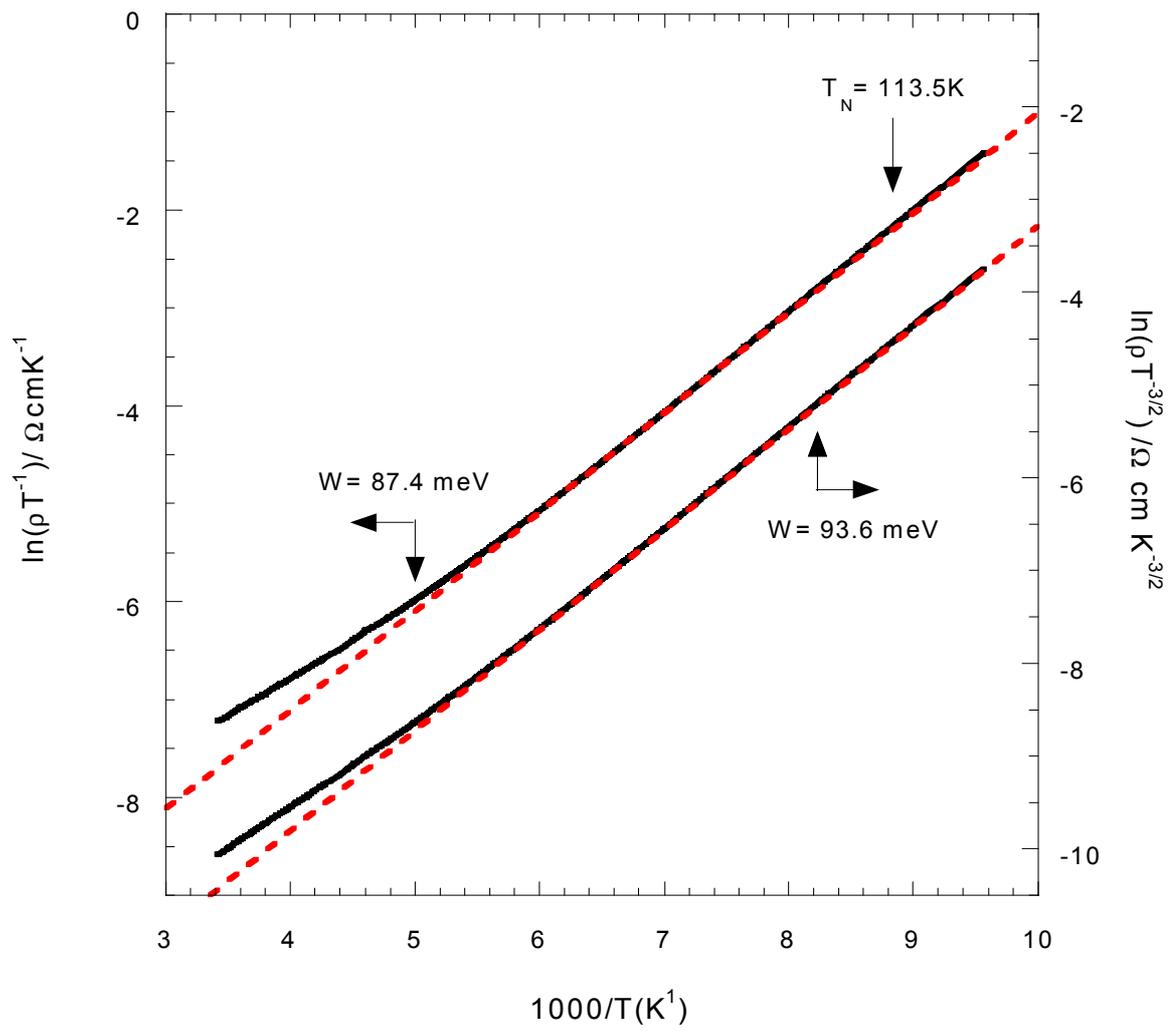

Figure 2 (Lago et al.)

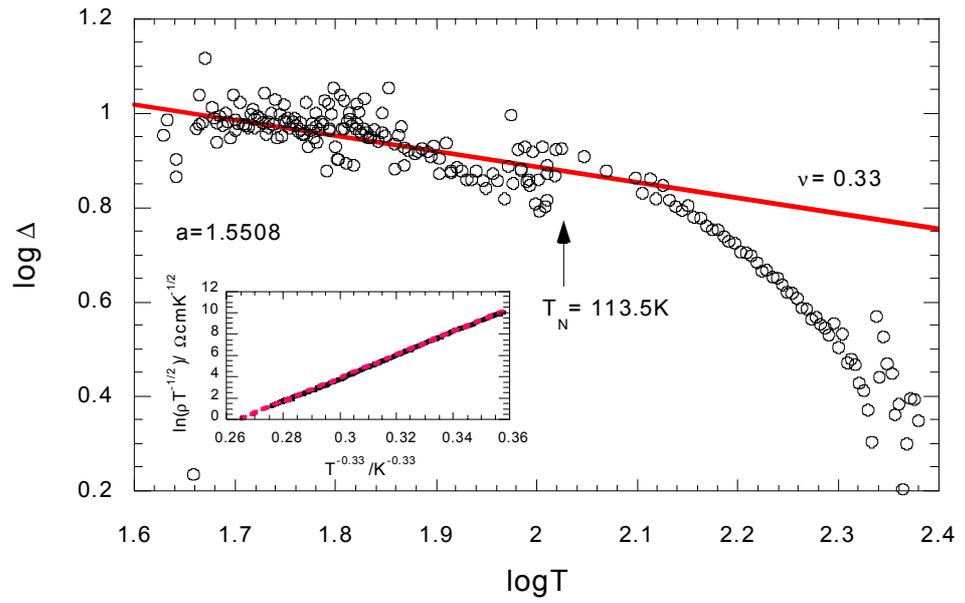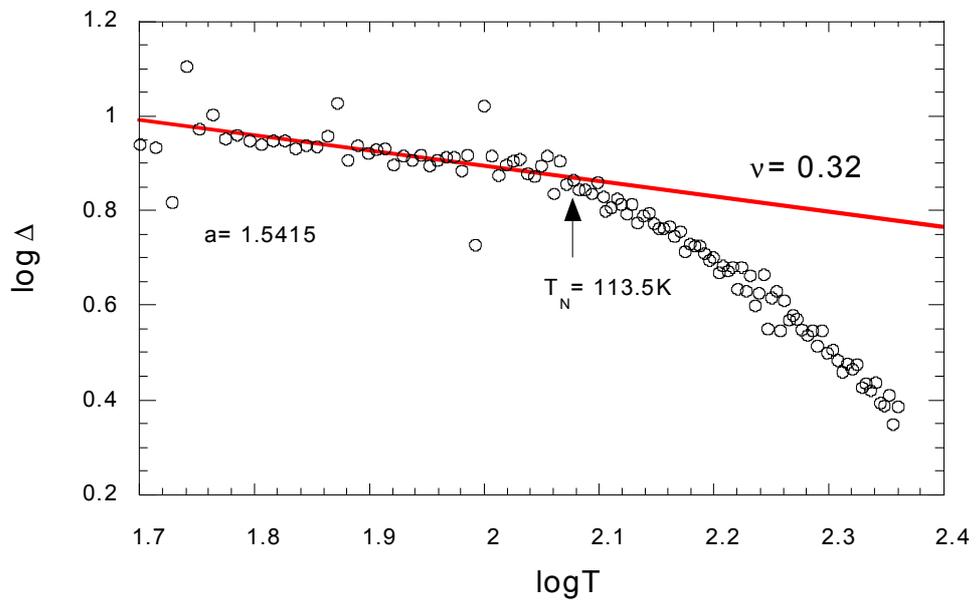

Figure 3 (Lago et al.)

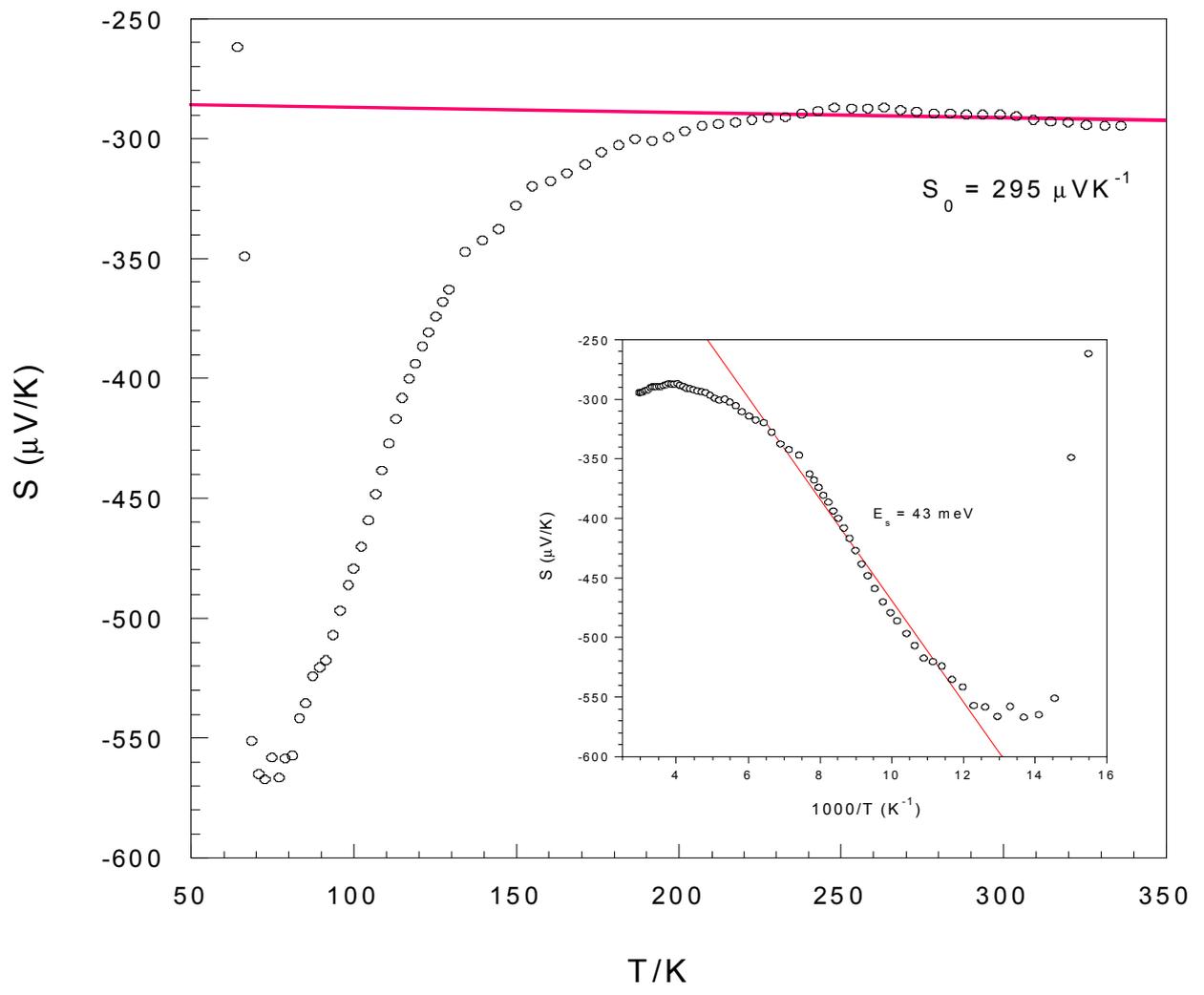

Figure 4 (Lago et al.)

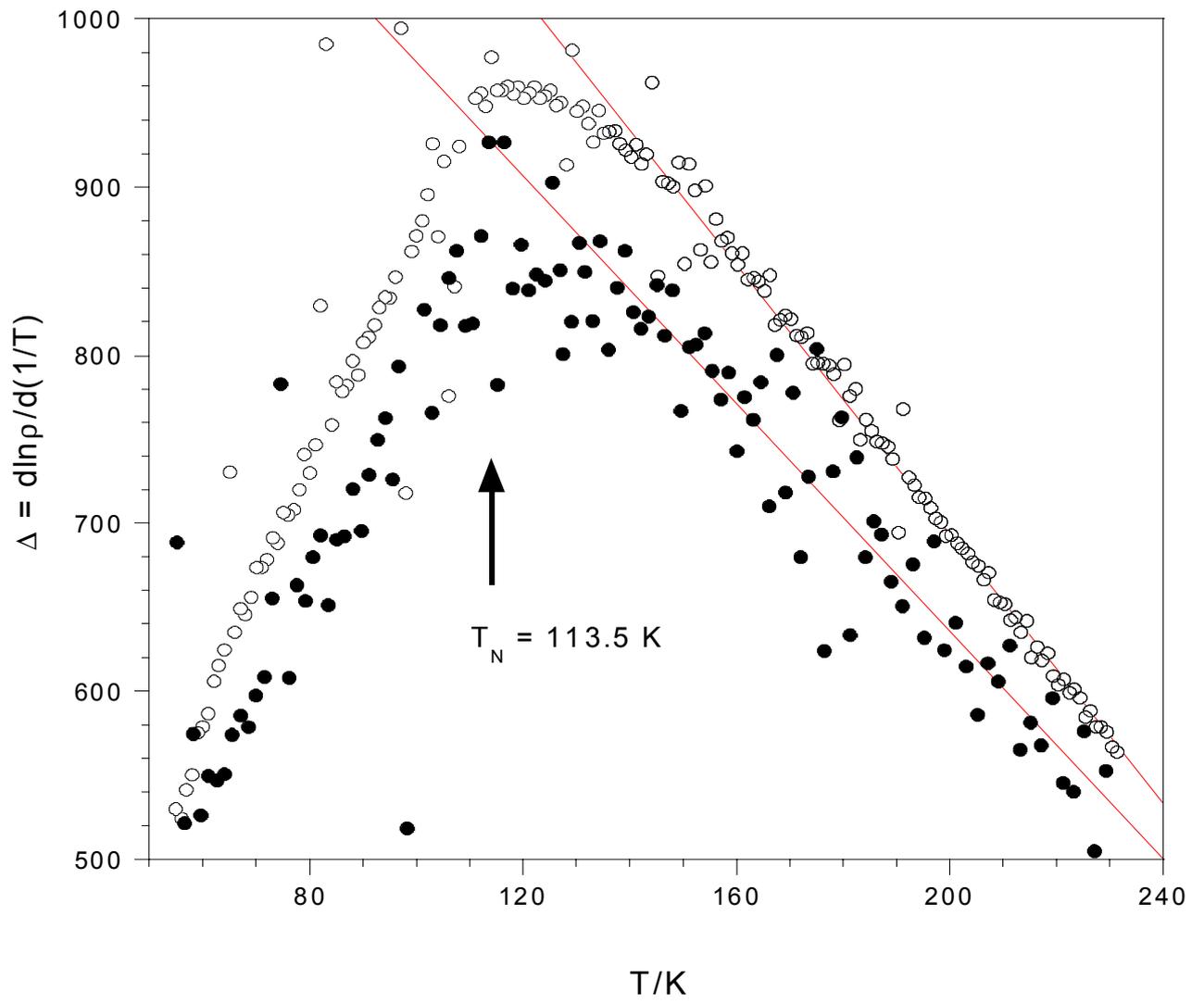

Figure 5 (Lago et al.)

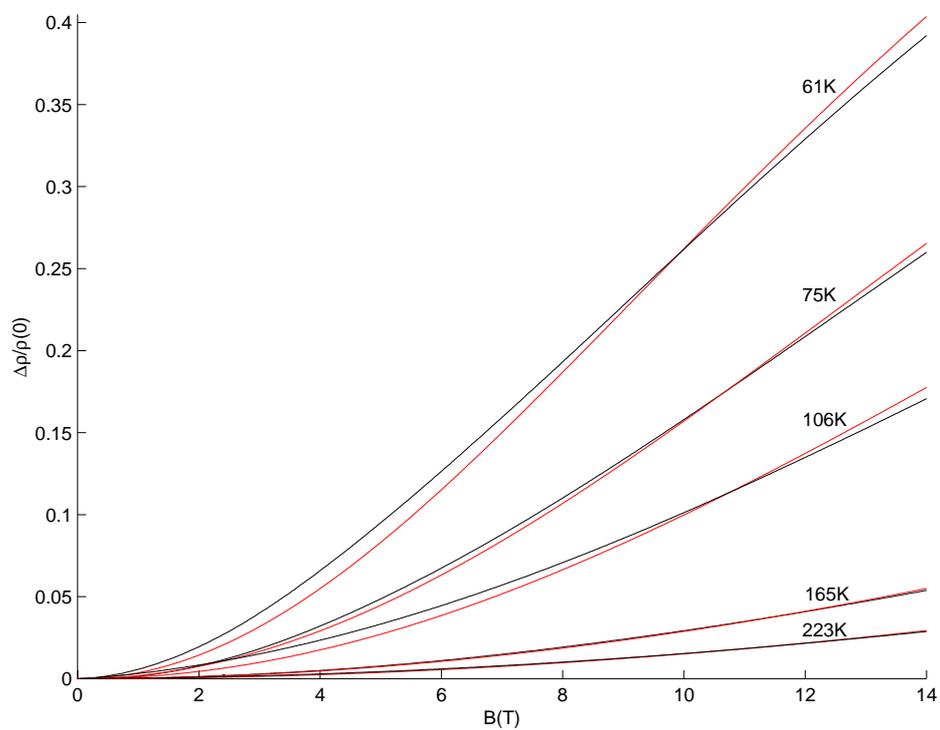

Figure 6 (Lago et al.)

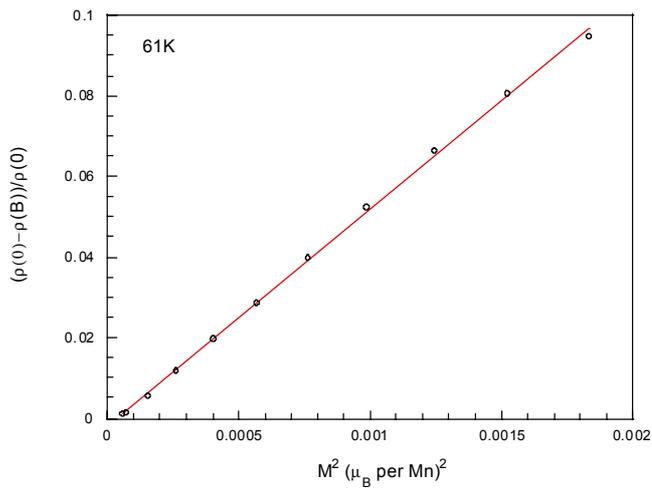
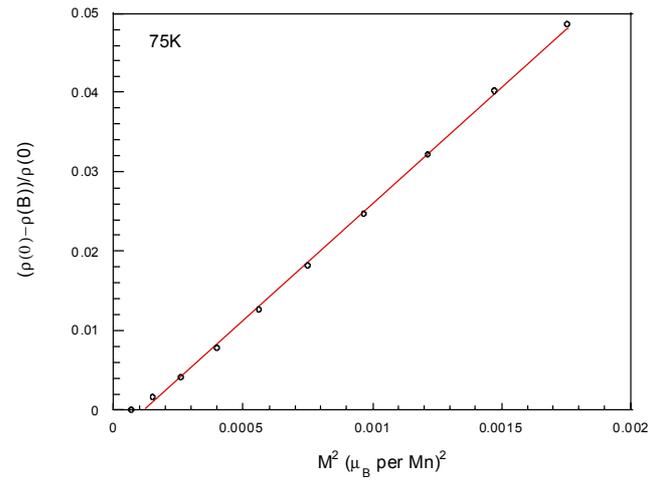
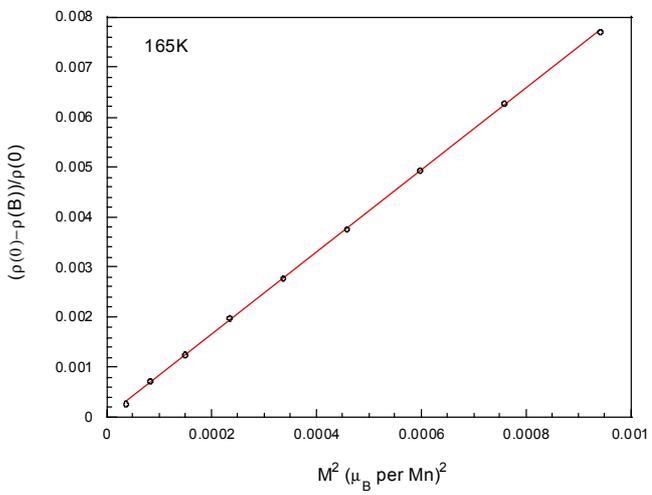
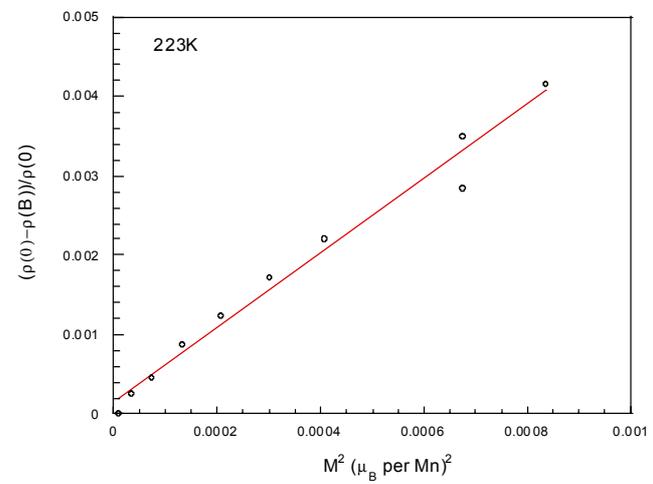

Figure 7 (Lago et al.)